# Filtering Signal Processes in Molecular Multimedia Memristors

ZhiYong Wang[1]†, LaiYuan Wang[1]†, Masaru Nagai[2], Linghai Xie[1]*, Haifeng Ling[1], Qi Li[3], Ying Zhu[1], Tengfei Li[1], Mingdong Yi[1]*, Naien Shi[1], Wei Huang[1,2]*

[1]Center of Molecular Solid and Organic Devices (CMSOD), Key Laboratory for Organic Electronics and Information Displays & Institute of Advanced Materials (IAM), Jiangsu National Synergetic Innovation Center for Advanced Materials (SICAM), Nanjing University of Posts & Telecommunications, 9 Wenyuan Road, Nanjing 210023, China.

[2]Key Laboratory of Flexible Electronics (KLOFE) & Institute of Advanced Materials (IAM), Jiangsu National Synergetic Innovation Center for Advanced Materials (SICAM), Nanjing Tech University (NanjingTech), 30 South Puzhu Road, Nanjing 211816, China.

[3]Physical Science Division, IBM Thomas J. Watson Research Center, 1101 Kitchawan Rd, Yorktown Heights, NY 10598, USA

*Correspondence to: iamlhxie@njupt.edu.cn, iammdyi@njupt.edu.cn and iamwhuang@njtech.edu.cn.

†These authors contributed equally to this work.

**Abstract**

To obtain precisely controllable, robust as well as reproduceable memristor for efficient neuromorphic computing still very challenging. Molecular tailoring aims at obtaining the much more flexibly tuning plasticity has recently generated significant interest as new paradigms toward the realization of novel memristor-based synapses. Herein, inspired by the deliberate oxygen transport carried by the hemoglobin in our blood circulation, we report a novel molecular-regulated electronic/ionic semiconducting (MEIS) platform ITO/MTPP/Al$_2$O$_{3-x}$/Al with a series of metallophorphyrins (MTPPs) to delicately regulate the ionic migration for robust molecular multimedia memristors. The stable pinched hysteresis resulted from the coordination-regulated ionic migration was verified by different device structures, operation modes, as well as the characterizations of scanning transmission electron microscopy with energy-dispersive X-ray spectroscopy (STEM-EDX) and X-ray photoelectron spectroscopy (XPS). Metal coordination-dependent device parameters such as potential and depression as well as retention curves further support the correlation between the coordination and stimulating flux-dependent memristive behaviors. In the 5,10,15,20-tetraphenyl-21H,23H-porphyrin zinc(II) (ZnTPP) synapse, we implement versatile emulations, mainly including transition from short-term memory (STM) to long-term-memory (LTM), learning experience and activity-dependent learning-memory process in integrated neuromorphic configurations based on the biological Hebbian rules, and develop the fresh Spike-Amplitude-Dependent-Plasticity (SADP) with the applications of signal filtering and habituation and sensitization which are beyond the prevalent Hebbian rules.

**Introduction**

Although the current prevalent computing architecture has sound memory hierarchy, including processor registers, SRAM caches, DRAM main memory, and virtual memory, the Von Neumann bottleneck of sequential processing still needs to be overcome in order to improve the computing speed and capacity as well as reduce energy consumption with regard to natural brains with the highly efficient neural network via the linkage of elementary synapses. Indeed, synaptic plasticity play a crucial role in learning, memory and forgetting as well as adaptation, together with thinking and emotion. As a result, artificial neuromorphic architectures based on memristors with the feature of bio-inspired plasticity become the next-generation solution for high-performance and intelligent computing. Plasticity of memristors endows the simple device with the controllable multiple states, history-dependent operation, and accumulative memory modes that are complementary to the bistable memory[1, 2]. Until now, various memristors have been extensively investigated in all the aspects ranging from theoretical frameworks, materials, detailed mechanisms as well as neuromorphic circuits[3]. Diverse synaptic plasticity such as spiking-rate-dependent-plasticity (SRDP) and spiking-time-dependent-plasticity (STDP) have been emulated by specific programming electrical parameters[4], including pulse amplitude, pulse number, pulse interval, as well as pulse width in metal oxide[5], inorganic chalcogenides[1], and organic memristors[6]. However, challenges still exist in terms of precisely control, robust as well as repeatability. Among various memristors, mechanisms of metal oxide memristors have been deeply explored, including electrochemical metallization[7], oxygen vacancies-migration-based valence change[8], drift and diffusion mechanism[9, 10]. The configuration of metal oxide memristors offers an ideal platform with in-situ oxygen source to provide the ionic migration. There are no report on the heterojunction memristors of organic multimedia and oxygen source layer that would offer more freedom for ionic migration to improve memristive performances with the intrinsically ideal mechanism of organic multimedia, although some organic materials were explored with the limited mechanism analysis[11]. Also, synaptic weight saturation is a severe problem that lead memory process to failure and the ability of sufficient learning-memory events[12, 13], although sometimes saturation has the special function for the homeostasis of neural networks[14]. It is emergent to improve the memristive capacity for ideal synaptic modulation[15].

The mixed ionic/electronic conducting (MIEC) is the ideal physical mechanism to construct two-terminal memristor with a typical pinched hysteresis[9]. In fact, the ionic migration is the essence reason to modulate the hysteretic conductive behaviors. Our idea is to find specific organic molecules serving as the high-performance multimedia of MEIC, compatible very well with the metal oxide memristor, to construct the heterojunction memristors. In the MIEC system, the current is given by the carrier processes, and the overall conductive state is dependent on the oxygen motion which transports through hopping process including binding and dissociation in each procedure. In the red blood cells of animals, the hemoglobin has the unique function of the revisable binding and delivery with oxygen for continuous oxygen transport.[16, 17] Inspired by the elaborate oxygen transport carried by the hemoglobin (iron-containing metalloprotein), the key material—MTPP is designed for the organic multimedia with the delicate management of oxygen transport. Here, we fabricate the preliminary devices ITO/MTPP/Al$_2$O$_{3-x}$/Al to demonstrate the novel molecular-regulated MIEC and the deliberate coordination-modulated effects of core metal center (Zn, Co, Ni, Fe) as well as their applications of versatile emulations, mainly including transition from STM to LTM, learning experience and activity-dependent learning-memory process in integrated neuromorphic configurations based on the biological Hebbian rules, and develop the fresh SADP with the applications of signal filtering and habituation and sensitization which are beyond the prevalent Hebbian rules.[18]

**Memristive switching behaviors in ZnTPP-based samples.** Fig. 1a shows the chemical structures of MTPP, in which M= Zn, Ni, Co and Fe-Cl refer to ZnTPP, NiTPP, CoTPP and FeTPPCl, respectively. To implement the potential dual-channel transport property of MTPP prototype, we first arbitrarily adopt ZnTPP serving as a model to fabricate the device with ITO/ZnTPP (~ 25 nm)/Al$_2$O$_{3-x}$ (~ 7 nm)/Al heterojunction structure (Fig. 1b) and have an area of 100 μm × 100 μm. Generally, ZnTPP is considered an organic p-type semiconductor material. According to the energy level diagram in Fig. 1c, the device ITO/ZnTPP/Al$_2$O$_{3-x}$/Al has a preference for hole transport due to a relatively smaller energy barrier between ITO/ZnTPP ($\Phi_B$ = 0.3 eV) (Schottky-like contact) relative to the larger barrier height between ZnTPP/Al$_2$O$_{3-x}$/Al ($\Phi_B$ = 1.9 eV). Therefore, ZnTPP layer acts as an ionic/electronic (mixed) conductor, where the resistive switching mechanism dominated by the ions distribution within the ZnTPP. Moreover, as a current limiter and oxygen reservoir, Al$_2$O$_{3-x}$ tunnel barrier also plays a key role for suppressing the metallization of electrodes.[19, 20]

For electrical testing, the electrical signals are applied to ITO electrode and Al electrode is grounded for all measurements in this study. As expected, the initial *I-V* curve of the device in its virgin state exhibits switching loop traversed as figure-of-eights, a typical switching hysteresis of memristor with rectifying characteristic (Fig. 1d). The ON/OFF current ratio measured at a bias of 7 V is close to 5. Then, detailed electrical characterization of these devices are further investigated. The switching endurance of at least 100 cycles is depicted in the inset of Fig. 1d. Although the switching loops after the first scan suffer from undesirable attenuation especially in negative direction, this ZnTPP device still shows a good stability without any saltation and inspired our enthusiasm to further study. The memristive behaviors of enhancing and suppressing hysteresis loops which are generally used to demonstrate the multiple resistive states of the device are employed to validate the ZnTPP device belong to memristor. As shown in Fig. 1e, during the 10 cyclic positive and negative voltage sweeps, the ZnTPP devices exhibit smooth, homogeneous change in resistance.[21] The ultimate current of ten positive and negative cyclic sweeps range from about 270 μA to 778 μA and -80 μA to -10 μA, respectively.

It should be noted that in this ZnTPP device an electroforming step is eliminated before the operation, thereby revealing that the device is determined by a well-defined non-filamentary mechanism.[9] To verify that the $O^{2-}$ migration accounting for the memristive switching in the ZnTPP memristor device, we change the scanning voltage range, bias scanning speeds, the device areas as well as device operating environments without package. Firstly, a series of devices were prepared with different areas ranging from $5\times10^3$ μm$^2$ to $5\times10^5$ μm$^2$ (Fig. S1). And we found that the current levels of ZnTPP device at HRS (increase from ~74 μA to ~1500 μA) and LRS (increase from ~454 μA to ~4048 μA) linearly scale with the device area, which confirm a homogeneous conduction mechanism rather than local filamentary conduction. As evidences of the memorization of stimulus-dependent ionic migration, the scanning rate-dependent memristive behaviors are investigated and the cyclic voltage sweeps with different strides ranging from 0.01 V/s to 5 V/s are applied into the ZnTPP memristors as shown in Fig. S2. We found that during positive and negative sweeps both the overall conductive level and the hysteresis area (from ~54.9 VμAcm$^{-2}$ to 2.0 VμAcm$^{-2}$) reduce monotonically as the scanning pace increases, as typical fingerprints of the charge-controlled memristor, which can be attributed to the improvement of total flowing $O^{2-}$ during one slow period. Moreover, the memristive behavior of rising hysteresis loops energized by 15 cyclic positive voltage sweeps at various rate patterns are presented in Fig. S3. In principle, during positive periodic scans, the augment of conductance gradually slacken, and the hysteresis area diminishes continuously, as shown in Fig. S4 for the width of oxygen-rich layer reaching limitation tardily. Instead, the hysteresis area and increasing current stride expand monotonically as the voltage-sweep rates decreases. These results strongly indicate that under the identical amplitude, lower voltage-sweep rates cause stronger stimulation during one stage. Thus, higher concentrations and mobility of $O^{2-}$ will certainly cause larger hysteresis area and current expand more rapidly.

The ZnTPP memristor devices exhibit excellent environmental suitability, which are benefit for the simulations and applications. As shown in Fig. S6, even though the ultimate current levels of ten positive cyclic sweeps ranges from 980 μA to 1396 μA are higher than their fresh counterparts, the ZnTPP devices after one-year storage still shows a satisfactory wider dynamic range of a tunable device conductance. On this basis, we found that the moist has certain influences on the device performance, mainly the promotion of device conductive level. It was observed that the current levels at 10 V and -10 V, respectively, increase from 1100 μA to 2800 μA and 18 μA to 80 μA with different moist ranging from 50% to 80% (Fig. S7). This result can be explained by that the exchange ability of $O^{2-}$ through ZnTPP matrix can be accelerated in condition that $H_2O$ coordinates with ZnTPP molecule (Fig. S8). After then, in order to verify the essential role of the oxygen source, the device without $Al_2O_{3-x}$ was fabricated. Our results suggest that the fresh device with a configuration of ITO/ZnTPP/Al exhibits the diode feature without producing the obvious hysteresis (Fig. S9a). The plausible reason is that the device only transports electrons/holes rather than $O^{2-}$. Further observation was made after exposing to atmosphere for two weeks (Fig. S9b). Interestingly, the device can also exhibits memristive behaviors, which is probably ascribed to that oxygen permeates into the device and then absorption into ZnTPP layers to serve as the mobile ions.

Furthermore, as evidence of ionic migration for memorization events, a sufficiently large voltage is necessary, because low voltage cannot promote the ions transport effectively (Fig. S6). We found that the *I-V* curve under low voltage scanning shows an obvious reversal current hysteresis including hysteresis direction and variation trend (Fig. 1f). Therefore, the devices exhibit a special sweeping voltage-dependent feature, namely, potentiating at larger voltage stimulations and depressing below the critical voltage. This sweeping voltage-dependent feature is analogous to the Matthew Effect, a profound selection mechanism of general evolution, which can be summarized as 'the strong get stronger and the weak get weaker'.[22] This adaptive performance in an elementary memristor enlightens us to extend fresh functions of the synaptic platform beyond the prevalently emulated synaptic rules for pursuing versatile information processing and effective neuromorphic configuration.

**Microscopic characterization of oxygen ion migration mechanism.** In order to directly confirm the memristive mechanism, STEM-EDX measurements of the device before and after electrical stimulations for 1000 s under 7 V were performed in a cross-section of ZnTPP film to probe the distribution of $O^{2-}$.[9] As shown in Fig. 2a, local conductive filaments cannot be found in organic layer after the electrical stimuli and the $Al_2O_{3-x}$ layer becomes thinner (decrease to ~5 nm) (Fig. S10), corresponding to the EDX elemental mapping images which clearly reveal that $O^{2-}$ is segregated from the $Al_2O_{3-x}$ layer.[23] Clearly, the peak of oxygen content ($C$) is enhanced ($C_0/C_{+7}$ = 148.0/168.8) and moves front about 1.5 nm after bias stimulating (Fig. 2b). In comparison, as expected, the Al profile is kept consistent before and after applied electric fields, thereby eliminating Al dominant memristive switching (Fig. S11). In addition, the oxygen concentration is remarkably enriched in the bulk organic layer, either the ZnTPP/ITO interface ($C_0/C_{+7}$ = 48.5/134.4) or the $Al_2O_{3-x}$/ZnTPP boundary ($C_0/C_{+7}$ = 51.5/61.4). The consecutively distributed oxygen profiles indicate the electric field-energized smooth and enduring ionic migration process, which can inherently ensure that the device conductivity changes continuously.

Furthermore, XPS was used to confirm that the oxygen species is $O^{2-}$ and explain the interaction between ZnTPP molecule and $O^{2-}$. First, the surface of prepared film was checked by XPS after clearly eliminating the Al as well as $Al_2O_{3-x}$ covering layers on the surface of ZnTPP film utilizing a stripping process (Fig. S12). The core-level XPS of Zn 2p and O 1s are presented in Fig. 2c-d. For XPS analysis, the peak energies were calibrated by assigning the major C 1s peak at 284.6 eV.[24] As can be seen in Fig. 2c, in comparison with the initial state, the distinct asymmetric O 1s spectrum was deconvoluted into two component peaks located at 530.85 and 531.21 eV, ascribed to charged oxygen species i.e., $O^{2-}$, $O^-$ in the Zn-O bonds, $O_x^-$ ions in the oxygen deficient regions.[25, 26] These results further verify that the $O^{2-}$ arise from $Al_2O_{3-x}$ layer enter into ZnTPP matrix after electrical polarization. In addition, the O 1s spectrum of the sample exposed in ambient atmosphere can be resolved into two main components as shown in Fig. 2c. And the two distinct peaks at 530.85 and 533.0 eV in the O 1s spectrum represent the incorporation of Zn-O and H-O-H components, respectively, which implies that few $H_2O$ molecule bonds with the ZnTPP matrix, corresponding to the increasing current as the moisture is raised as discussed above, because the coordination between ZnTPP and $H_2O$ can reduce the binding energy of migrated $O^{2-}$. These qualitative components obtained by curve fitting are consistent with the reported values.

In addition, the Zn 2p spectrum presents the same full width at half-maximum, namely, chemically equivalent (Fig. 2d). The only difference is a very small binding energy shift of 100 meV for the Zn 2p spectrum when compared with the pristine state, indicating that in the stimulated system abundant $O^{2-}$ are free to assemble, and furthermore we ascribe the binding energy shift to the minimization of the interaction energy with a small amount of charge reorganization between $Zn^{2+}$ and $O^{2-}$. This demonstrates that $Zn^{2+}$ in porphyrin ring plays the role in providing the binding sites for migrated $O^{2-}$, and, the sharing of electrons from the donor of $O^{2-}$ leads to an electrostatic binding interaction, namely the coordination bond, which results in the small binding energy shift of the Zn 2p spectrum. Thus, based on this, we were able to identify the presence of coordination bond between $Zn^{2+}$ and $O^{2-}$. In conclusion, the XPS results together with the STEM-EDX characterizations confirmed the resultant smooth memristive behaviors for the delicate coordination-regulated ionic migration.

According to the overall memristive operations and characterizations, the memristive mechanisms are proposed. As described in Fig. 2e, it is depicted the coordination sites and potential movement paths for the migration of $O^{2-}$ controlled by the local ZnTPP molecules. When applying the enough positive voltage, the redox process for the $Al_2O_{3-x}$ layer leads to the generation of $O^{2-}$. Then, these negatively charged ions will release from oxygen resevoir and bind with ZnTPP molecules instantly (Fig. S13). The binding energies between $Zn^{2+}$ and $O^{2-}$ is about -1.866 eV, which is inadequate to forbid the migration of $O^{2-}$ toward next molecules. Therefore, $O^{2-}$ suffer from binding and dissociation processes reversibly. In comparison with the uncontrollable random migration pathway in oxide devices,[27] the regulation of this novel coordination process can confine the $O^{2-}$ successive translocation along next ZnTPP molecules under electrical fields. The resultant redistribution of the $O^{2-}$ will affect essential interfacial parameters at the $Al_2O_{3-x}$/ZnTPP and the ZnTPP/ITO interface simultaneously, especially for Schottky-like interface which contributes to the resistive switching.[28] As a result, the electronic transport, namely, the device resistance, will be changed. In the opposite case, the $O^{2-}$ reversely move towards Al electrode until reenter into $Al_2O_{3-x}$ layer via self-limiting oxidation which will produce the NDR performance.[29] Therefore, the initial ion distribution recovered and leads to the unsymmetrical gradual resistance change memristive characteristics.

As mentioned, excepting the resultant excitatory trends of intense stimuli, the overall inhibitory trends gradually increase under stronger weak stimuli, which we ascribe to the larger polarization of ZnTPP domains (Fig. S14).[30] In principle, the weak external electric field cannot break the coordination bond effectively, and there exist the redistribution of bonded $O^{2-}$, which results in the inversely enhanced polarization inside the active layer. And the polarization is expected to be more distinct under relatively stronger weak electric field. Eventually the polarization of inner polar groups leads to the enhancement of the localized internal electrical field and induces degressive conductive state, as indicated by the current expression $I \propto (\overrightarrow{E_{ext}} - \overrightarrow{E_{in}})$.

**Metal-dependent plasticity.** In our coordination modulated memristive prototype, the organic multimedia plays the key role in robust coordination-assisted $O^{2-}$ transport and dynamic regulation for memristive behaviors, which closely relates to the binding affinity and dissociation between ionic $O^{2-}$ and the MTPP. Therefore, the device deserves to be sensitive to different MTPP active layers characterize by different $E_{M-O}$.

To deeply get insight into the exquisite mechanism of the coordination modulated memristive behaviors, we detailedly investigate the coordination bonding effect on the ionic transport by utilizing different MTPP mediums including ZnTPP, NiTPP, CoTPP and FeTPPCl. A series of memristors are fabricated with identical configurations as well as uniform thickness of deposited MTPP layers to examine the device performance.

Fig. 3 shows the measured device current (blue curves) as a function of the scanning time under ten cyclic positive (0 ~ 10 V) and negative voltage sweeps (0 ~ -10 V). In contrast to ZnTPP memristor, the ultimate current of ten positive cyclic sweeps ranges from about 612 μA to 1190 μA in NiTPP memristor (Fig. 3b) and 777 μA to 1480 μA in CoTPP memristor (Fig. 3c), respectively. And during the negative sweeping, the current continuously decreases together with a NDR peak of -1351 μA at -6.7 V in ZnTPP memristor, -558 μA at -3.7 V in NiTPP memristor, and -240 μA at -3 V in CoTPP memristor (Fig. S15), respectively. The results reveal that the overall change of current level gradually increases and NDR peak voltage gradually decreases in ZnTPP, NiTPP and CoTPP memristor.[31]

To deeply understand the delicate differences of the memristor performance and coordination between $O^{2-}$ and three MTPP materials, the metal-oxygen bond length and binding energies are calculated through the method of density functional calculations. We have noticed that the imparity of metal ligand characterized by different binding affinity and dissociation constants between metal ligand and $O^{2-}$ has significant effect on the memristive behaviors, including the overall conductive level and current increasement under the identical measurements. The metal-oxygen bond length gradually prolongs and the binding energies gradually attenuates for ZnTPP, NiTPP and CoTPP molecules (**Table. S2**), corresponding to the increasing overall conductivity, which is ascribed to $O^{2-}$ are relatively more difficult to diastasis the bondage of coordination bond into the oxygen-poor region, according to the kinetic equation:

$$v = v_0 exp\left[-\left(E_{M-O} - \frac{1}{2}\varepsilon qa\right)/k_B T\right]$$

Where $v$ is the escape frequency under the external electrical field $\varepsilon$, and $v_0$ is the attempt-to-escape frequency. $a$ denotes the average migration length.[32] Moreover, in these memristors with regard to the NDR performance related to the ionic back motion, the gradually reduced NDR peak voltage shown in Fig. S15 is in accordance with the decreasing difficulty degree of ionic back motion, and further demonstrates that for the MTPP medium with weak $E_{M-O}$ it is easier for the $O^{2-}$ to reenter the $Al_2O_{3-x}$ layer in order of the bond energy ($E_{Co-O} < E_{Ni-O} < E_{Zn-O}$).[33]

In order to demonstrate the crucial impact of coordination-assisted ionic migration channel, the FeTPPCl is selected as a comparison deliberately which is characterized by a bonded chlorine atom with the iron ligand on one side of the planar organic molecule. As shown in Fig.3d, significantly, the FeTPPCl device performs differently from the three aforementioned memristors with relatively rough *I-V* curves, high operation voltage (~14 V) as well as low current level (about 8~16% of other MTPP devices), which can be attributed to the fact that the bonded chlorine inhabiting the metal center can decrease the positions for bonding $O^{2-}$ inside the FeTPPCl matrix, thereby leading to dismal oxygen permeability liable for the poor properties. Thus these measurements further indicate that the migration of $O^{2-}$ are enslaved to the coordination bond and can be significantly influenced by the environment around the coordination center.

As other evidence to demonstrate the specific ionic migration channel in the advanced MTPP memristors, we furthermore make the control experiment of device platforms constructed by 5,10,15,20-tetraphenyl-21H,23H-porphyrin (TPP) without metal ligand. In comparison with the MTPP devices, during positive voltage sweeps TPP elements exhibit a large current level together with flat pinched hysteresis (especially the existence of overlap and even anticlockwise hysteresis between forward and back scans during 8 V ~ 10 V) (Fig. S15e). The following scanning current exhibits an increasing trend with the similar shape. And during the negative cyclic scanning, the current gradually reduces, which is characterized by the unobvious NDR profile of the first cycle. From the memristive curves, we consider that the $O^{2-}$ can be transported inside TPP layer during about 4 V ~ 8 V, and after 8 V the overlapping curves can be ascribed to the intense accumulation of randomly migrated $O^{2-}$. Inside the TPP matrix without the delicate management of coordination bond, the $O^{2-}$ are transported crudely through the interspace of TPP molecules at a high speed as indicated by the resultant higher conductive level (memorized ionic migration in the wide zone). And the unobvious NDR profile during negative sweeps (no need of sufficient operation voltage to promote ionic back motion) indicates that the back-diffusion speed is relatively fast at the absence of coordination regulation. The discussed results and the following durative characteristics under fixed bias in Fig. 3f indicate that the organic TPP materials lacking of metal ligand is not capable of subtly modulating the successive ionic migration, and the conductivity of TPP device will rapidly increases to a peak value as discussed in the following section, namely a non-monotone curve. This platform provides a further verification of the key strategy—coordination-assisted ionic transport for finely controlled memristive behaviors.

The coordination-assisted $O^{2-}$ migration suggests that the device conductance can be persistently adjusted by inputting pulse sequence with fixed bias, namely the electric-pulse-induced resistance (EPIR) performance,[34, 35] which is considered as the basic response to investigate the complex stimulus-response rules for synaptic imitations.[36] After the multiple cyclic scanning, we also make comparisons between the serials devices focusing on the time-dependent conductive variation utilizing a sequence of electrical pulses with fixed voltage amplitude of 10 V and retention time of 10000 s for a correlation between the memorization characteristics and the binding affinity of MTPP (Fig.3g). The conductance of CoTPP device rapidly increases with a maximum steep after ~100 s retention, and it rapidly reaches to the saturation value, especially 95% of the saturation (~718 μS) at ~860 s. For NiTPP device, the conductance also rapidly increases although with a deceleration at ~1000 s, especially 95% of the saturation (~415 μS) at ~2000 s. In comparison, the conductance of ZnTPP device increases at a relative slow speed. The saturation value is not attained even after 10000 s retention, and the conductance increases to 95% of the final value (~518 μS) at ~4390 s, which indicates the excellent anti-saturation ability under continuous electrical pulses. Besides, in the FeTPPCl device the enhanced hindrance restricts the ionic migration, which results in the sluggish increase at the initial stage, overall low conductivity as well as roughly durative characteristics. And in the TPP device, as expected, the conductive state rapidly increases to a peak value (~622.5 μS) at ~65 s with a slope larger than CoTPP device then rapidly reduces to a low conductive level. In conclusion, the long-term smooth, gradually changing retentive characteristics of ZnTPP memristor presented in Fig. 1 also demonstrate the effectiveness of the coordination-assisted ionic migration. And furthermore, the slower migration of $O^{2-}$ in ZnTPP for higher $E_{M-O}$ verifies the significant role of coordination band in modulating the memristive behaviors as discussed above. Herein, given its remarkable memristive behaviors, namely long-term smooth, gradually changing resistance adjustment, preferable anti-saturation properties as well as low energy consumption potential, ZnTPP is utilized specially in our subsequent research to implement the versatile neuromorphic functions.

**Mathew effect for signal filtering.** The ZnTPP memristor with impressive memristive behaviors, smooth *I-V* characteristics and multiple intermediate states for improving the modulatory space, provides us a fine prototype to simulate the synaptic functions.[3, 11, 37, 38] Biologically, in our brain the memory refers to the capability to retain external stimulations, based on the physiological synaptic weight with different strength and durations.[1, 39] Generally, two types of synaptic plasticity, STP and LTP, and their transitions dominate our memory events. The higher-level memorization, namely transitions from STM to LTM, can be achieved through the shaping process under larger doses (longer term or more times) of learning. These familiar events are significant for the learning and memory activities in our brain.[40] And achievements of the adjustment of synaptic memory is of prevalent interest, which are based on the fundamental Hebbian rules implying that stronger accumulated activations result in stronger memorized responses. The bio-inspired functions are also effectively implemented in ZnTPP-based artificial synapse relied on the coordination-regulated ionic migration, including the transition from STM to LTM (Fig. S17, 18), learning experience-enhanced memory (Fig. S19), which indicate that the memory level can be enhanced by longer term learning actions and more learning periods. Notably, although under the long-term repeated stimulations with different intensities, the multi-stage learning and memory responses still keep smooth and gradually changing characteristics, demonstrating the superiority of the coordination-assisted ionic migration and consolidation for neuromorphic emulations. Furthermore, as an intuitive, concrete expression in our visual neural system, we can utilize the advanced electrical performances in integrated configurations composed by ZnTPP synapse pixels to implement the stimulus activity-dependent learning and memory behaviors (Fig. S24). This indicates that the versatile elements can function as self-adjustive hardware visual systems and generate the adaptive learning and memory pattern effectively.[1]

Presently, significant learning and memory processes based on the adjustable neuroplasticity in the frameworks of fundamental Hebbian rules and the revised version STDP rules have been implemented extensively through the memristive devices.[10, 31, 41, 42] However, compared with the sophisticated biological regulation process in realistic neural networks (RNNs), the developed imitations are preliminary forms which lack some essential regulatory properties and sophisticated features.[43-45] Actually, in a synapse, the ability to remodel its response mode dynamically according to external variable stimuli for self-adjusting weight and memory is the key capacity, termed as the activity-dependent property.[46-48] And it is likewise for the neuromorphic device. It is desirable to endow the sophisticated characteristics with the emulated neural activities for more effective neuromorphic configurations. With regard to the ZnTPP memristor system, in addition to the activity-dependent learning and memory events, more importantly, the ZnTPP-based sensory neural configuration can implement the signal filtering and memorization based on the inherent property of stimulus-dependent mode selection.

In the ZnTPP artificial synapse, except for the increasing trend under intense stimuli satisfying the additive property in Hebbian frameworks, there exist an intriguing performance that the device response tends to decrease gradually under weak stimuli (Fig. S25).[1, 49] This intelligent performance in ZnTPP memristor enlightens us to implement the signal filtering and memorization for pursuing adaptive neuromorphic configuration. As shown in Fig. S27, the activity-dependent information filtering and memorizing processes are visualized via two contrastive integrated displays with 8 × 8 pixels in each display, and each pixel is implemented by a single ZnTPP artificial synapse. As a prevalently-emulated memory case, a sequence of stimulations (Fig. S26) is applied into the memristor pixels corresponding to the graphic word 'Hi' and the obtained impression is continuously increased from ~30.7 μS to ~66.7 μS as a function of the stimulus amount. However, generally, there exists the residual impression of memorizing other signals a moment ago, analogous to the biological characteristics of persistence of vision. In this situation with residual background information (the average conductance level is ~30.4 μS) (Fig. 4), the signals of word 'Hi' are continuously input into the memristor pixels, and the impression level is enhanced gradually. While the background information stored in the memristor pixels outside the 'Hi' image spontaneously relaxes as soon as the stimulations are removed (decrease to ~20.0 μS after the 30th input). This operation is analogous to the common emulation case, where the ZnTPP sample play the role as a general memristor.[1] Biologically, the identification effect of a pattern is dependent on the contrast ratio of the pattern rather than its simplex intensity, which determines the effectiveness of memorizing the object. For the neuromorphic hardware-implemented architecture, as discussed above, the ZnTPP-based memristor can select different modes of information processing in different stimulating conditions, namely enhancement mode under strong signals and attenuation mode under weak signals. The signals with high and low intensities (Fig. S28, 29), which corresponds to the input signals derived from the image and the surrounding, respectively, are applied into the memristive arrays. As a result, the impression of the word 'Hi' is progressively deepened from ~40.1 μS to ~66.5 μS, while the surrounding background information is continuously weakened to ~6.9 μS, and thus the contrast ratio of the word is enlarged as more pulses are applied (Fig. 4b). Consequently, the ZnTPP memristor with matthew effect contributes to the potential application in adaptive information filtering and memorizing, and extends the fresh functions successfully for the neuromorphic hardware.

We furthermore utilize the activity-dependent pattern adjustment in the ZnTPP memristor to emulate the brain-inspired signal processing beyond the fundamental Hebbian rules, that is, habituation and sensitization.[50-54] To this end, we prepare three integrated displays including 8 × 8 pixels to visualize the information processing of activity-dependent mode selection. As shown in Fig. S31 and Fig. S32, three images of letter 'I', 'A' and 'M' are stimulated by alternate modest stimuli ($V$ = 5 V, $W$ = 100 ms, as habituation signals) and intense stimuli ($V$ = 10 V, $W$ = 100 ms, as sensitization signals) with different intervals ($T_I$ = 1.5 s, $T_A$ = 7.5 s, $T_M$ = 15 s), which are followed by a series of low voltage pulses ($V$ = 3 V) to monitor the varying device state, respectively. With regard to the letter 'I', the device conductive state gradually decreases from ~2.8 μS to ~1.2 μS as the moderate pulses are applied, implying that it is increasingly difficult to activate this artificial synapse as more periods are applied, namely the occurrence of neural habituation. And the conductive level gradually increases from ~5.5 μS to ~11.1 μS as the intense stimuli are applied, resembling the sensitization. Furthermore, the suppressed conductive level is also enhanced from ~6.5 μS to ~3.4 μS through the sensitive signals. As expect, the sensitized response still gradually reduces as soon as the moderate stimuli are applied. The overall response at the second habituation stage is also enhanced by the sensitive activations, as the neural sensitization implies that sensitized response is characterized by a universal response enhancement commonly under the whole range of stimuli, including strong and moderate ones.

By comparison, for the input-dependent mode selection of the letters 'A' and 'M' with prolonged intervals of habituation and sensitization inputs. During the initial habituation process, the overall synaptic response reduces more as the interval extends, as 'A' ($H_1$: ~1.3/0.6 μS, $S_1$: ~1.7/5.0 μS, $H_2$: ~2.8/1.2 μS) and 'M' ($H_1$: ~0.3/0.1 μS, $S_1$: ~0.4/2.3 μS, $H_2$: ~1.5/0.5 μS), which conforms to the SRDP learning rules. Under the periodic habituation stimulations with extended intervals, the following monitoring pulses of 3 V can suppress the recovery of (and even restart) the habituation responses, which results in the gradually weakened response (Fig. S33). And during the following stimulations under alternant intense and moderate pulses, the changing tendency is real-timely controlled by and selectively subject to the input signals with different intensities, thereby indicating the robust characteristics of the activity-dependent mode selection. From the pictorial activity-dependent habituation and sensitization processes of three letters 'I', 'A', and 'M', it can be conclude that as the interval is prolonged the reached conductive level is reduced, especially the imperceptible letter 'M' for both the strengthened habituation effects as well as the weakened sensitization effects (Fig. S34). The aforementioned input-dependent habituation and sensitization responses lay a solid foundation for adaptive signal selection, namely paying less attention to unimportant signals and more attention to novel stimuli.[52] Thus the ZnTPP-based synaptic device provides a prototype for practical implementation of versatile neuromorphic configurations.

In summary, we proposed a concept of organic multimedia on site-assisted molecular dual transport as a robust solution to fabricate the artificial synaptic devices. The microscopic nature of both the $O^{2-}$ migrating and the coordination binding is revealed for the first time by in situ STEM-EDX and XPS analysis, and supported by the method of density functional calculations and the correlativity between metal types and *I-V* behaviors. As a result, the MTPP that can be regarded as the state-of-the-art concept model of organic multifunctional electron/ion semiconductors to fabricate the drift-type memristors with a well-defined device configuration of ITO/MTPP/$Al_2O_{3-x}$/Al, in which $Al_2O_{3-x}$ play a role of ion source and MTPP serve as the ionic transporting media. Moreover, various scanning modes or stimulation pulse sequences have been explored to emulate multiple synaptic functions, such as STP and LTP (STM/LTM), SRDP, 'learning-experience' behaviors, habituation and sensitization. In particular, the capability of emulating Matthew Effect directly at the device level enable the implement of signal filtering just rely on memristor devices, which contribute to decreasing the number of active elements in future computing systems. We hope that our results will guide the molecular design rules with the feature of reliability, repeatability and anti-saturation, as well as pave an encouraging pathway toward practical implementation of artificial neural networks for neuromorphic computing.

**Acknowledgments:** The project was supported by the National Basic Research Program of China (2014CB648300, 2015CB932200), National Natural Science Foundation of China (61475074, 21274064, 61136003), National Natural Science Funds for Excellent Young Scholar (21322402), Changjiang Scholars and Innovative

Research Team in University (IRT_15R37), the Natural Science Foundation of the Education Committee of Jiangsu Province (14KJB510027), Natural Science Foundation of Jiangsu Province (BK20160088), the Qing Lan Project of Jiangsu Province, the Innovation Team of Talents in Six Fields of Jiangsu Province (XCL-CXTD-009), Synergetic Innovation Center for Organic Electronics and Information Displays, Excellent Science and Technology Innovation Team of Jiangsu Higher Education Institutions (2013), A Project funded by the Priority Academic Program Development of Jiangsu Higher Education Institutions (PAPD,YX03001).

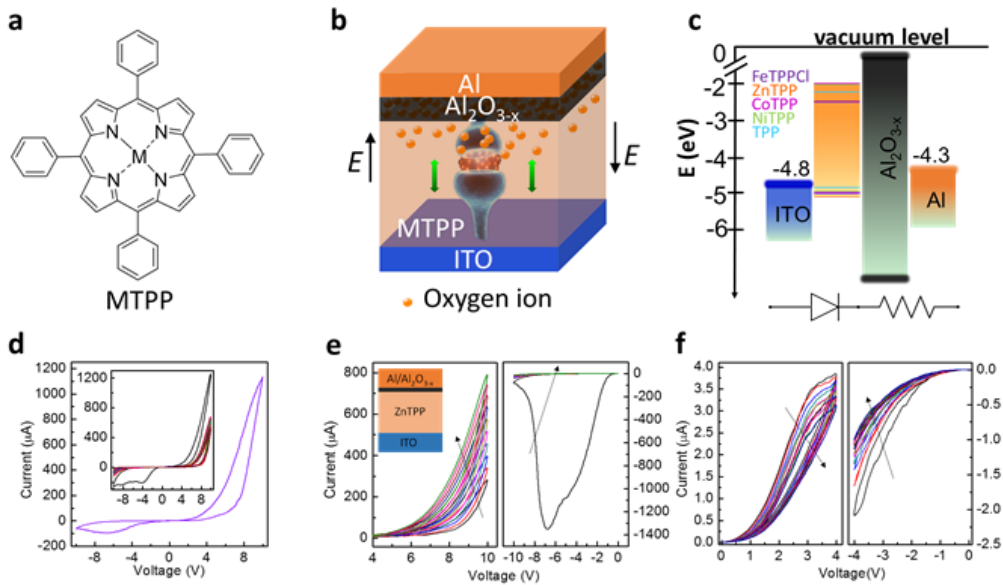

**Fig. 1| Memristive behaviors of ZnTPP device. a**, Chemical structures of MTPP. M = Zn, Ni, Co and Fe-Cl represent of ZnTPP, 5,10,15,20-tetraphenyl-21H,23H-porphyrin nickel(II) (NiTPP), 5,10,15,20-tetraphenyl-21H,23H-porphyrin cobalt(II) (CoTPP), 5,10,15,20-tetraphenyl-21H,23H-porphyrin iron(III) chloride (FeTPPCl), respectively. **b**, Schematic representation of ITO/ZnTPP/$Al_2O_{3-x}$/Al-stacked structure. The movement of internal $O^{2-}$ is used to model the switching. **c**, Energy-level diagram of pristine device. The energy values of the HOMO and LUMO levels of each MTPP molecular that we have probed are listed in table (**Table. S1**). **d**, Typical counter figure-eight hysteretic loop current-voltage (*I-V*) characteristics of a ZnTPP memristor with the moderate thickness of $Al_2O_{3-x}$ layer. The inset shows stable hysteretic bipolar memristive switching characteristic at a constant sweep rate of 0.1 Vs$^{-1}$ for 100 consecutive cycles. **e**, Hysteretic *I-V* loops of ZnTPP devices measured at 10 V with voltage-sweep rates of 0.1 Vs$^{-1}$. The device schematic is shown in the inset. **f**, Hysteretic *I-V* loops of ZnTPP devices measured at 4 V with voltage-sweep rates of 0.1 Vs$^{-1}$. The sweeping voltage-dependent feature is analogous to the Matthew Effect.

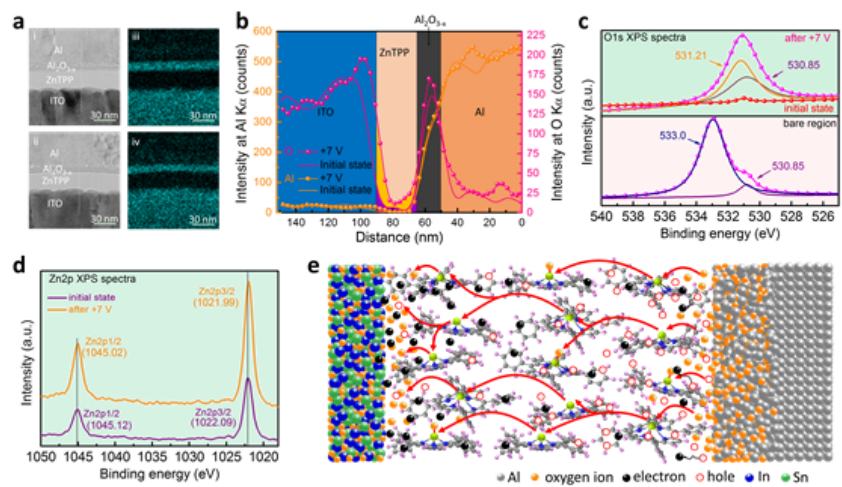

**Fig. 2| STEM-EDX and XPS analysis of migratory $O^{2-}$. a**, STEM images and the

corresponding EDX elemental mapping of O (blue) of the ZnTPP film in initial state (i and iii) and after stimulation at 7 V (ii and iv). **b**, EDX line profiles of O (pink) and Al (orange) scanned from the Al to the ITO side along the white dashed lines in the STEM images in **a** (i and ii). The EDX line profiles reveals that the oxygen concentration is remarkably enriched near the ZnTPP/ITO boundary, and other regions are slightly increased. The line profiles of O are obtained by subtracting the background noise determined from the Al TE. **c**, O1s XPS data collected from pristine state, after electrode polarization and bare region. The four colored lines except pink indicate Gaussian fits, with peaks at 530.85, 531.21, 532.83 eV and 533.0 eV. **d**, Zn2p XPS data collected from pristine state and after electrode polarization. In the polarized films, the measured spectrum (orange line) shift 100 meV when compared with the pristine state. **e**. Illustration of $O^{2-}$ translocation along a chain of ZnTPP molecules, as postulated to occur for the memristor mechanism. The mobile routes of $O^{2-}$ that move along the ZnTPP chain are labelled in red.

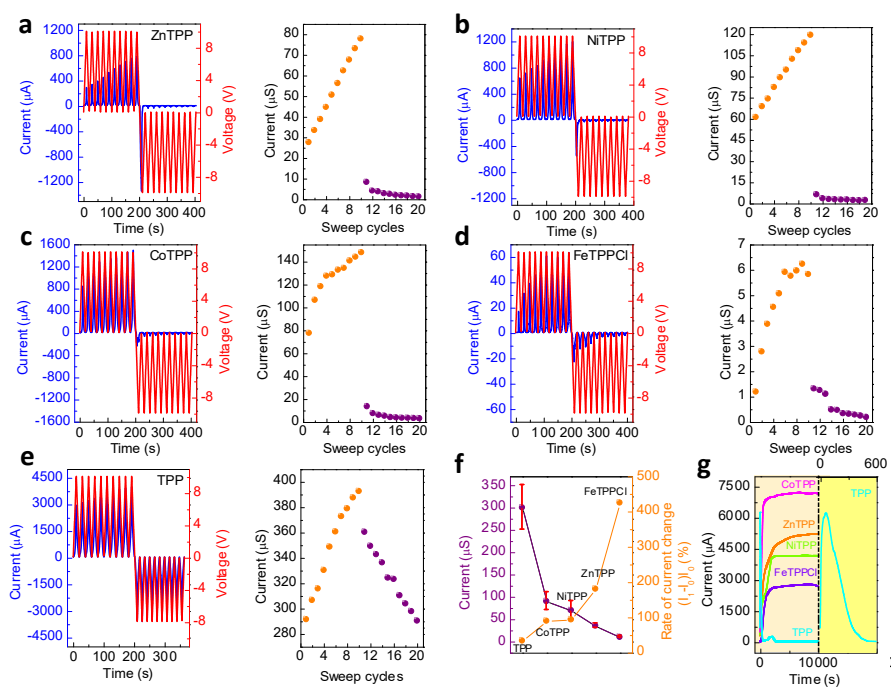

**Fig. 3| Memristive behaviors of disparate MTPP-based devices. a-e**, Variable current and voltage versus time, and corresponding conductivity of MTPP thin films with identical configuration, which are plotted from the data in Fig. S15. **f**, Variable conductivity of MTPP after first scan (violet) and rate of current change after 10 sweep. **g**, Anti-saturation behaviors of disparate MTPP-based devices.

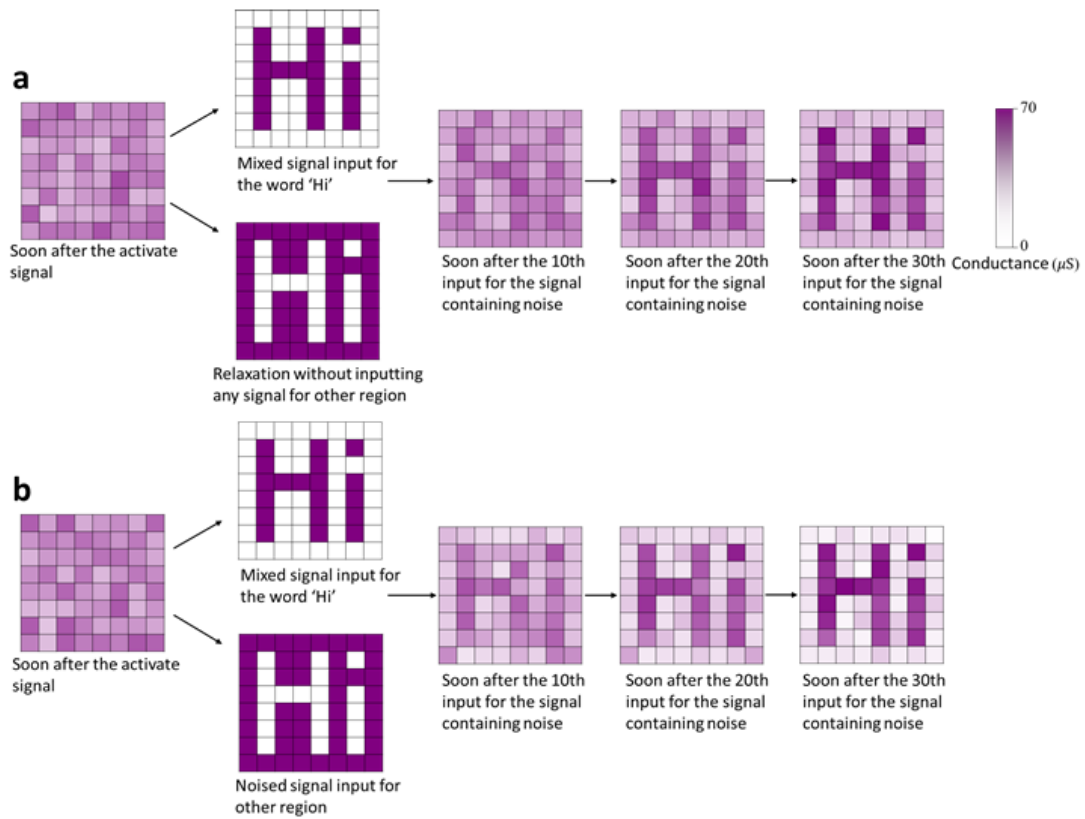

**Fig. 4| Visualizations of activity-dependent information filtering and memorizing processes. a**, The signals of word 'Hi' are continuously input into the memristor pixels with residual background information (the average conductance level is ~30.4 μS). Subsequently, the impression level of the word 'Hi' is enhanced gradually as a function of the stimulus amount, meanwhile, the background information outside the 'Hi' image spontaneously relaxes to ~20.0 μS after the 30th input. **b**, Two types signals with high and low intensities, which represent the input signals derived from the image and the surrounding (i.e., signal and noise), respectively, are applied into the memristive arrays. Significantly, the contrast ratio of the word 'Hi' is enlarged as more pulses are applied. The ratio of the average conductance level of the word and the surrounding background information is about ten soon after the 30th input, which is thrice as high as the Fig.4a.

# Supplementary Materials for

# Filtering Signal Processes in Molecular Multimedia Memristors

ZhiYong Wang[1]†, LaiYuan Wang[1]†, Masaru Nagai[2], Linghai Xie[1]*, Haifeng Ling[1], Qi Li[3], Ying Zhu[1], Tengfei Li[1], Mingdong Yi[1]*, Naien Shi[1], Wei Huang[1,2]*

[1]Center of Molecular Solid and Organic Devices (CMSOD), Key Laboratory for Organic Electronics and Information Displays & Institute of Advanced Materials (IAM), Jiangsu National Synergetic Innovation Center for Advanced Materials (SICAM), Nanjing University of Posts & Telecommunications, 9 Wenyuan Road, Nanjing 210023, China.

[2]Key Laboratory of Flexible Electronics (KLOFE) & Institute of Advanced Materials (IAM), Jiangsu National Synergetic Innovation Center for Advanced Materials (SICAM), Nanjing Tech University (NanjingTech), 30 South Puzhu Road, Nanjing 211816, China.

[3]Physical Science Division, IBM Thomas J. Watson Research Center, 1101 Kitchawan Rd, Yorktown Heights, NY 10598, USA

*Correspondence to: iamlhxie@njupt.edu.cn, iammdyi@njupt.edu.cn and iamwhuang@njtech.edu.cn.

## Methods

Device fabrication

The devices were fabricated on commercial ITO-coated 3×3 cm glass substrates (Kaivo Optoelectronic Technology corporation) with sheet resistance about 10 Ω/sq. The etched ITO work as bottom electrode, and the area of the synaptic device was 0.01 $mm^{-2}$, defined by 0.1-mm-wide ITO stripes overcrossed by 0.1-mm-wide top Al stripes. The substrates are washed by sequential procedures in an ultrasonic bath with acetone, ethanol and deionized water for 10 min in each procedure and then dried in the vacuum oven at 120℃ for 30 minutes. The commercial MTPP (molecules produced by Sigma Aldrich corporation) without further purification and Al as well as $Al_2O_{3-x}$ layers were deposited by thermal vacuum evaporation deposition at $\sim 10^{-5}$ Torr. MTPP films with ~25-nm-thick are first grown using the thermal vacuum method with a evaporation growth rate of ~0.3 Å/s. The thickness of the films was detected by employing Step Profiler and verify by using STEM. Then, the $Al_2O_{3-x}$ films were fabricated at $\sim 10^{-4}$ Torr by a slow evaporation method with a rate of 0.1~0.3 Å/s, ensuring that Al can be spontaneous oxidized.[1, 2] This process results in an $Al_2O_{3-x}$ layer (~ 7 nm) characterized by a high resistivity media, as verified by the STEM and EDX analysis. The $Al_2O_{3-x}$ layer used for the junctions was ~7 nm thick, as verified by the STEM and EDX analysis, and play a key function of oxygen reservoir and suppresses any chemical reaction between the Al electrode and the active layer. Finally, a top Al electrode (about 125 nm) is intentionally defined utilizing conventional evaporation process (0.5~1.5 Å/s) by using a shadow mask (to obtain strips of Al), for purpose of forming an Ohmic contact at the metal/oxide interface via generating a large amount of oxygen vacancies.[3]

Electrical Measurements

The electrical characteristics with purpose-designed profiles are measured using a Keithley4200 semiconductor parameter analyser with self-designed testing software. The top electrodes of the junctions were grounded during all electrical measurements, and we define a flow of current from bottom to top electrode as the positive bias. All the measurements were carried out under ambient conditions without any encapsulation. In Fig. 2a, to verify the migration of $O^{2-}$ by using STEM-EDX, a series of positive pulses ($V$ = 7 V, $W$ = 100 ms, $T$ = 100 ms) were applied to ensure that $O^{2-}$ can enter into the ZnTPP films. Fig. 3j presents the remarkable anti-saturation of MTPP based synaptic devices, where the devices were stimulated by $10^5$ consecutive pulses ($V$ =10 V, $W$ = 100 ms, $T$ = 100 ms). In Fig. S24, to carry out concrete psychological behavior in a neuromorphic visual system, the letters 'I', 'A', 'M' were severally stored three times using ten identical amplitude inputs ($V$ = 10 V, $W$ = 100ms) with different pulse intervals ($T_I$ = 1.5 s, $T_A$ = 7.5 s, $T_M$ = 15 s), and then the current is constantly monitored by the read pulses ($V$ = 3 V, $W$ = 100 ms, $T$ = 100 ms) immediately after the last stimulus in the series. The biological behavior of habituation and sensitization were dramatically simulated in Fig. S32. Alternate modest stimuli ($V$ = 5 V, $W$= 0.5 s) and intense stimuli ($V$ = 10 V, $W$ = 0.5 s) were used to imitate habituation and sensitization, respectively, and the current levels (monitored by the read pulses, $V$ = 3 V, $W$ = 100 ms, $T$ = 100 ms ) are visualized in images of the letters 'I', 'A', and 'M' corresponds to identical stimuli except intervals ($T_I$ = 1.5 s, $T_A$ = 7.5 s, $T_M$ = 15 s).

Computational details

Our density functional calculations were carried out using the Quantum Espresso (QE) package[4] with norm-conserving pseudopotentials and generalized gradient approximation (GGA) exchange-correlation functionals parameterized by Perdew-Burke-Enzerhof (PBE).[5] The initial molecular structures were taken as the tetrabenzoporphyrin molecules in Ref.[6]. We simulated the 'single porphyrin molecule in vacuum' geometry by constructing a simple tetragonal unit cell centered by the metal ions of the MTPP molecule. The size of the unit cell is selected so that the spacing between the molecules of adjacent images is larger than 25 Bohr in all directions. The total energy of the systems are tested to converge into a $10^{-5}$ Rydberg window by using this unit cell. A convergence criterion of $10^{-7}$ Rydberg was used for all the self-consistency calculations. The molecular structures are relaxed until all the force components were less than $10^{-4}$ Rydberg/bohr. All the calculations were performed at Γ-point. We studied $O^{2-}$ adsorption mechanism on the metal ion of ZnTPP, NiTPP and CoTPP molecules. The $O^{2-}$ binding energies were calculated by

$$\Delta E_O = E(molecule + O) - E(molecule) - \frac{1}{2}E(O_2)$$

where E(molecule+O), E(molecule) and E($O^2$) are the total energies of the surfaces with one $O^{2-}$ adsorbed, the porphyrin molecule, and the gas phase oxygen molecule, respectively. The binding energies, metal-oxygen bond length, and the Gibbs free energy of the $O^{2-}$ adsorption on the three different molecules are listed in **Table. S2**. The relaxed structures are shown in Fig. S16.

STEM-EDX analysis.

The ultrathin sample was prepared by the focused ion beam (Hitachi FB-2100). STEM-EDX analysis was performed with JEM-ARM200F scanning transmission electron microscope and JED-2300T analysis station operating at 200 kV. The line profiles of oxygen was corrected by subtracting the background noises determined from the Al electrode whereas aluminum profile without any processing.

X-ray photoelectron spectroscopy.

The TE with 1000 μm diameter was polarized by sweeping the voltage in the range of 0 to 7V at a sweep rate of 0.1 Vs$^{-1}$ in order to inject $O^{2-}$ into ZnTPP film. Then, XPS (PHI Quantera II) was performed on the bare surface of polarized films after removing the TE and $Al_2O_{3-x}$ layer which were precision stripped by using specific tape (Fig. 2b). The surface of prepared film was checked by XPS and eliminating the Al or $Al_2O_{3-x}$ contaminants on the surface of ZnTPP films after precision stripping process, and ensuring that the reliability of analysis results.

**Supplementary Text**

Synaptic functions simulation

Biologically, in our brain the memory refers to the capability to retain external stimulations, based on the physiological synaptic weight with different strength and durations.[7, 8] Generally, two types of synaptic plasticity, short-term plasticity (STP) and long-term plasticity (LTP), and their transitions dominate our memory events. The higher-level memorization, namely transitions from STM to LTM, can be achieved through the shaping process under larger doses (longer term or more times) of learning. These familiar events are significant for the learning and memory activities in our brain.

In order to investigate the memory consolidation through longer term stimulations, as shown in Fig. S20, five series of learning pulses ($N$ = 5, 10, 15, 20, 25, 30) with identical amplitude ($V$ = 10 V), duration ($W$ = 100 ms) and interval ($T$ = 100 ms) are applied into the ZnTPP-based memristor, respectively. And then a pulse sequence with low read voltage ($V$ = 1 V, $W$ = 100 ms, $T$ = 100 ms) is utilized to detect the relaxing current level as soon as the stimulations are removed, which is considered as the forgetting occurrence and can be described accurately by a stretched exponential decay equation:

$$I = (I_0 - I_\infty) \exp\left[-(\frac{t-t_0}{\tau})^\beta\right] + I_\infty$$

where $I_0$ and $I_\infty$ are the synaptic weights retained at the initial state and the stabilized state, respectively, and $\tau$ is relaxation time constant which is particularly utilized to evaluate the rate of changing response.

Obviously, after being stimulated by only 5 pulses, the synaptic weight gets almost inappreciable after 100 s (from 100% to 7.0 %). After more learning stimulations, more and more memory contents are remained (decline to 20.1 % for 30 pulses). Longer and longer relaxation time constants (increase from ~6.6 s to ~20 s after different stimulations ranging from 5 to 30) are extracted from the forgetting curves in Fig. 20, which indicate that the forgetting curve gradually tends to be more solid along with the increasing stimulations, namely transition from STM to LTM (Fig. S21). It is worthwhile to mention that benefiting from the regular ionic motion the forgetting curves in different devices are featured by smooth variations throughout each response phase.

Except for the longer term learning, more learning periods can also enhance the memory level in ZnTPP based artificial synapse. In this situation, the repeated learning stages are programmed for emulating the essential 'learning–forgetting–relearning' behaviors of biological learning activities to investigate the transition between STM and LTM.[8] To implement the 'learning–forgetting–relearning' process, intense stimuli ($V$ = 10 V, $W$ = 0.5 s, $T$ =0.5 s) and following read stimuli ($V$= 3 V, $W$ = 0.5 s, $T$ = 0.5 s) are intermittently applied into the memristor as shown in Fig. S22. At the initial learning stage, the conductive level continuously increases under 50 stimulations with a learning time constant $\tau$ of ~100 s, and then spontaneously decays (as long as $10^3$ s) in the absence of learning inputs. We set the ultimate current level after 50 pulses at the initial state as the standard of synaptic weight of 100%. And it is remained 30% after $10^3$ s, indicating that the activated current is relaxed to an intermediate state. Interestingly, the current state can be enhanced again at a higher speed as soon as the relearning signals are input (15 pulses are need to obtain the 100% memory level and the learning time constant $\tau$ = 90 s), and then spontaneously relaxes with 40% retention value after $10^3$ s, which is followed by one more relearning-forgetting process with a shorter τ of 80 s and 55% retention value during the forgetting process. For the three learning periods, the decreasing learning pulses and the reducing τ to obtain the same memory level indicate that it gets easier for subsequent relearning. Furthermore, for the three forgetting periods, higher and higher retentive memory and increasing forgetting time constant (~29.1 s, ~36.6 s and ~64.7 s) demonstrate the transformation tendency of STP-to-LTP. The promotion of memory consolidation through relearning process embodies the effectiveness of the synaptic 'learning-experience' emulations and strong accumulative memory effect in ZnTPP element (Fig. 23).

From the perspective of internal physical mechanism, the 'learning–experience' behaviors can be ascribed to the extended migration of $O^{2-}$ and more adequate separation of oxygen and aluminum, which further conforms to the rearranges of $O^{2-}$ inside the ZnTPP matrix. As labeled in Fig. S22a, initially, a series of electrical stimulations create an oxygen-rich region in the ZnTPP layer as described, and the boundary (violet flat) refers to the conduction front. And the forgetting behavior occurs due to the back-motion of relaxed $O^{2-}$ to bond with the interfacial aluminum afresh when the stimulations are removed, namely the partial retreat of the conduction front as shown in Fig. S22b. As evidence of the back-motion of driven $O^{2-}$, the resultant reverse current is detected under 0 V as soon as the stimulating pulses are removed (Fig. S24). And the reverse current becomes intenser after higher voltage stimulations for the larger scale motion of activated ions. The red flat in Fig. S22 is employed to represent the position of conduction front after the retracement. For the overall 'learning–forgetting–relearning' process, it can be deduced that the retreat distance is relative long at the initial forgetting stage and then gradually shortened at the subsequent forgetting processes for more adequate ionic separation, corresponding to the progressive consolidation of memory in ZnTPP synapse. Notably, although under the long-term repeated stimulations with different intensities, the multi-stage responses still keep smooth and gradually changing characteristics, demonstrating the superiority of the coordination-assisted ionic migration and consolidation for neuromorphic emulations.

As shown in Fig. S24, the activity-dependent memorization and forgetting processes are visualized via three integrated displays with 8 × 8 pixels in each display, and each pixel is implemented by a single ZnTPP artificial synapse. In this adaptive sensory system, to visualize the activity-dependent memory consolidation, images of three letters 'I', 'A', 'M', which are combined by a series of elements. These three letters are respectively stimulated three times utilizing ten identical inputs (V = 10 V, W = 100 ms) in each stage with different intervals ($T_I$ = 1.5 s, $T_A$ = 7.5 s, $T_M$ = 15 s) as learning signals, followed by a series of low voltage pulses (V = 3 V) to monitor the varying device state after resting for 20 s and 40 s. (Fig. S22). The representative change in the conductance of an individual synapse response to various intervals and stored periods are presented in Fig. S26.

With regard to the letter 'I', the memory level gradually increases to ~17.1 μS as the learning actions are operated continuously and then decays spontaneously over time (decrease to ~0.75 μS after 20 s and ~0.34 μS after 40 s). The relaxed memory can be enhanced again beyond the previous level during subsequent relearning process (increase to ~23.1 μS and ~27.6 μS after the second and third learn period), namely the learning-experience effect, which can consolidate the stored memory and suppress the memory decay. By comparison, for the learning-memory processes of the letters 'A' and 'M' characterized by weakened learning activities with gradually prolonged intervals, the memorization cannot reach to the level as high as the letter 'I' (~10.1 μS, ~16.9 μS and ~21.3 μS for the letter 'A'; ~3.0 μS, ~12.2 μS and 18.3 μS for the letter 'M'). And the overall conductance is degressive as the interval extends, conforming to the SRDP learning rules. Especially, after the first learning process, the memory of letter 'M' appears to be blurring after 20 s (~0.08 μS) and imperceptible after 40 s (~0.017 μS), corresponding to the STM performance. From the pictorial interval-dependent learning and memory processes, it can be concluded that the STP can be promoted to transform into the LTP after repeat learning as well as concentrated learning. For the robust synapse arrays, we emphasize that the memorizing and forgetting responses are characterized by gradual change benefitting from the regular ionic process, rather than rough or saltation, which is conductive to emulate the gentle accumulative process of learning-memory.

| MTPP | $\varepsilon_{HOMO}$ | $\varepsilon_{LUMO}$ |
|---|---|---|
| ZnTPP | -5.13 | -2.39 |
| NiTPP | -5.03 | -2.04 |
| CoTPP | -5.03 | -2.02 |
| TPP | -4.91 | -2.19 |
| FeTPPCl | -5.07 | -2.44 |

**Table. S1|** The energy values of the HOMO and LUMO levels of each MTPP molecular. Obtained by using theoretical calculation.

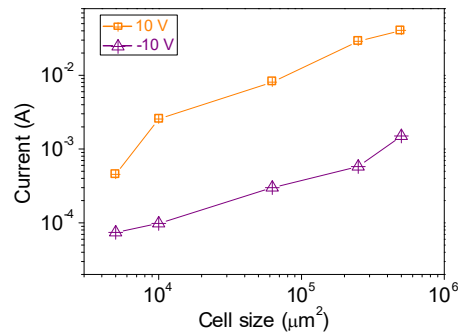

**Fig. S1|** HRS/LRS-current-value dependence on cell size ranging from $5\times10^3$ μm$^2$ to $5\times10^5$ μm$^2$ at -10V/10 V. The current value at each cell size is calculated by the averaging of measurement at 5 different devices.

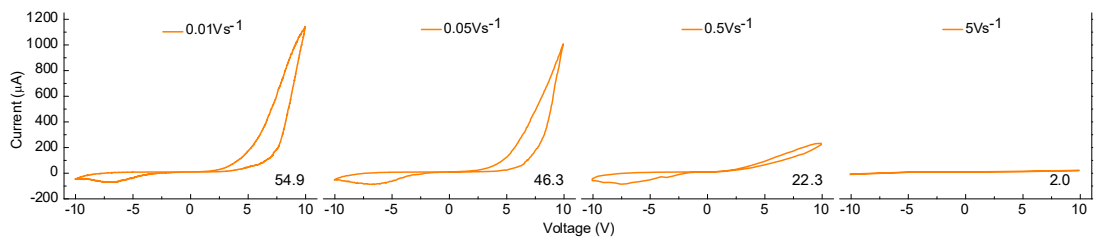

**Fig. S2|** Variable hysteretic *I-V* loops with various voltage-sweep rates in the range of 0.01-5 Vs$^{-1}$. The numbers in each graph represent the hysteretic area (VμAcm$^{-2}$) of the loop.

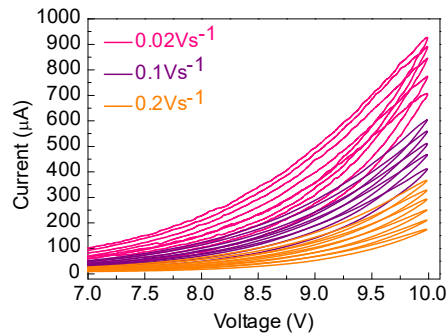

**Fig. S3|** *I-V* profiles for systematically decreased sweep rates to probe the time-dependence of the memristance. Current hysteresis curves at different scanning speeds (i.e., 0.02 Vs$^{-1}$, 0.1 Vs$^{-1}$ and 0.2 Vs$^{-1}$), and the hysteresis area and increasing current stride expand monotonically as the voltage-sweep rates decreases.

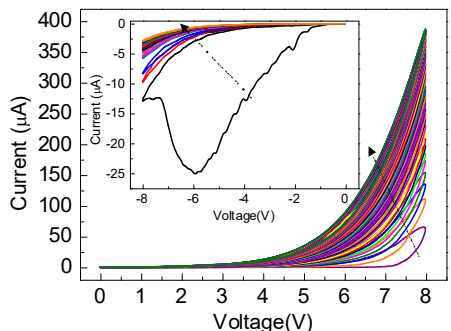

**Fig. S4| Typical *I-V* relationship of ZnTPP memristor during 100 positive periodic scans**. The augment of conductance gradually slacken, and the hysteresis area diminishes continuously, as fingerprints of the memristor. The inset shows the subsequent *I-V* relationship during negative periodic scans.

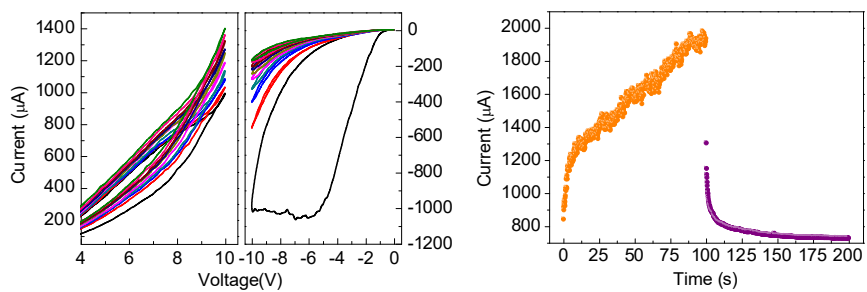

**Fig. S5|** Memristor properties of the ZnTPP device after store in ambient atmosphere one year.

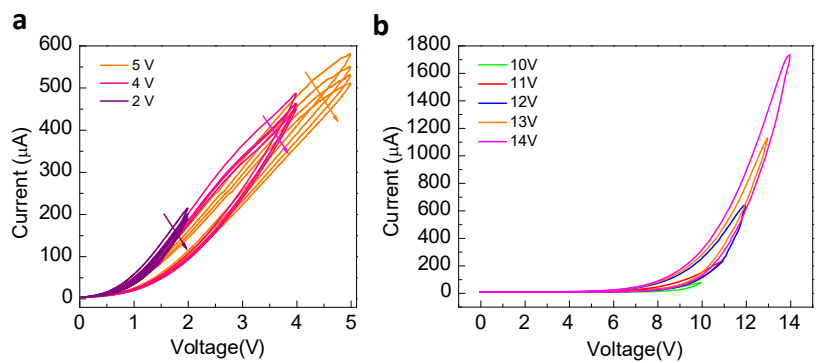

**Fig. S6| Sweeping voltage-dependent feature, which is analogous to the Matthew Effect. a**, Current hysteresis curves at different sweeping voltages (i.e., 2 V, 4 V and 5V), revealing that the current decreasing at the less than 5 V. Test after operation at 10 V in order to inject enough O$^{2-}$ into ZnTPP film. **b**, Tuning current curves with enhancive sweeping amplitudes, as a result of increasing hysteresis loops and hysteretic area.

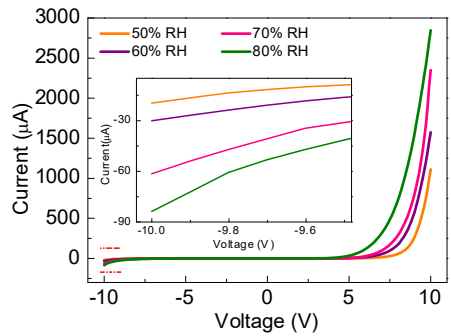

**Fig. S7**|The electrical response of ZnTPP films under different humidity, indicating that the current increases as the RH is raised from 50% to 80%.

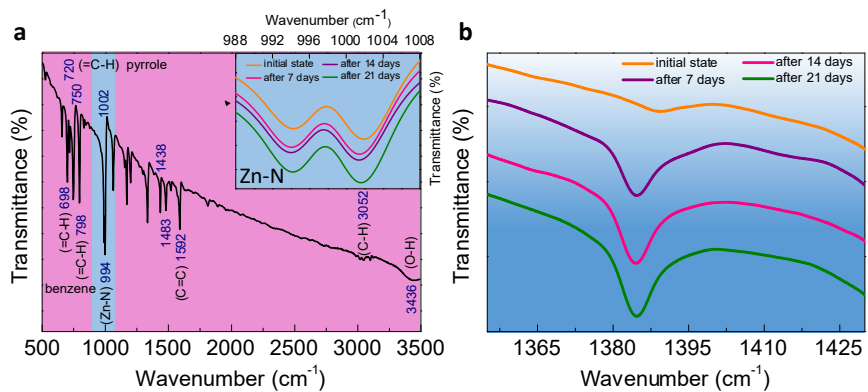

**Fig. S8| a**, FTIR spectra of ZnTPP powder obtained from fresh powder. The inset shows the Zn-N stretching vibrations after store in ambient atmosphere ranging from fresh to 21 days. **b**, The enhanced sharp peak around 1384 cm$^{-1}$ is attributed to H-O-H bending vibration, is assigned to a small amount of $H_2O$ in the ZnTPP matrix, which in agreement with that determined for humidity effect.

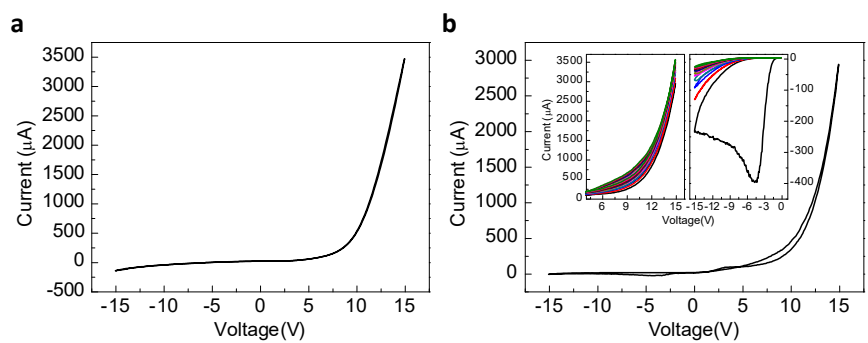

**Fig. S9|Device performance of ITO/ZnTPP/Al devices. a**, The fresh devices with a configuration of ITO/ZnTPP/Al could not produce the obvious hysteresis. **b**, After store in ambient atmosphere two weeks, the devices can also exhibit memristive behaviors.

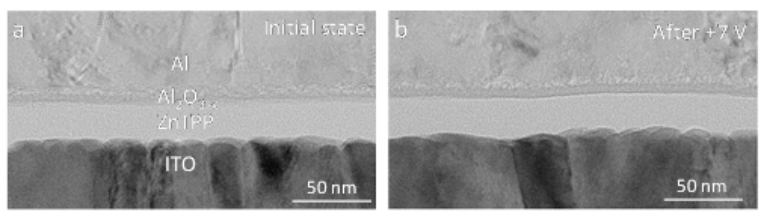

**Fig. S10|** STEM images of the ZnTPP film in initial state (**a**) and after stimulation at 7 V (**b**). Scale bar, 50 nm.

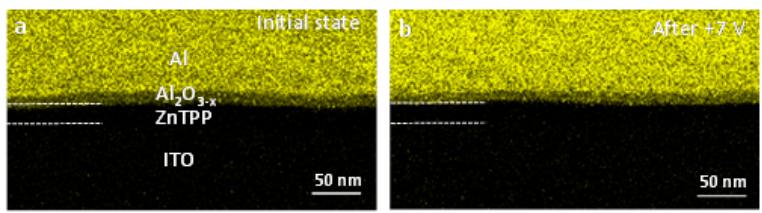

**Fig. S11|** EDX elemental mapping of Al (yellow) in initial state (**a**) and after stimulation at 7 V (**b**). Scale bar, 50 nm.

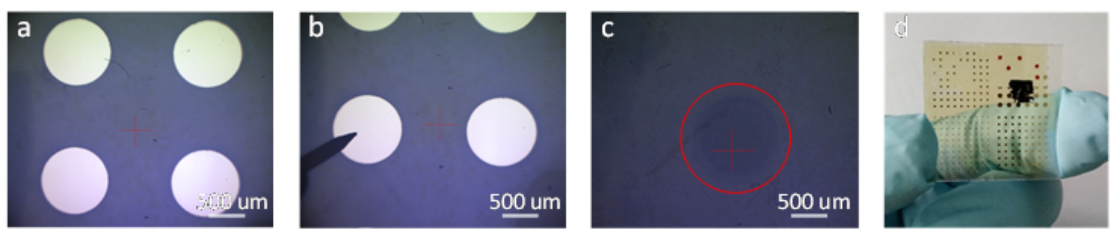

**Fig. S12| Procedure to prepare the sample for the XPS analysis. a,** Optical micrograph of the pristine device with 1000 μm diameter TE. **b,** The BE was polarized by sweeping the voltage in the range of 0 to 7V, and 0 to 7V followed by 0 to -7 V at a sweep rate of 0.1 Vs$^{-1}$ in order to create unequal O$^{2-}$ diffusion indwell ZnTPP films. **c,** After polarization, the TE were precision stripped by using specific tape. **d,** Optical image of the prepared sample, and the red dot were the test points.

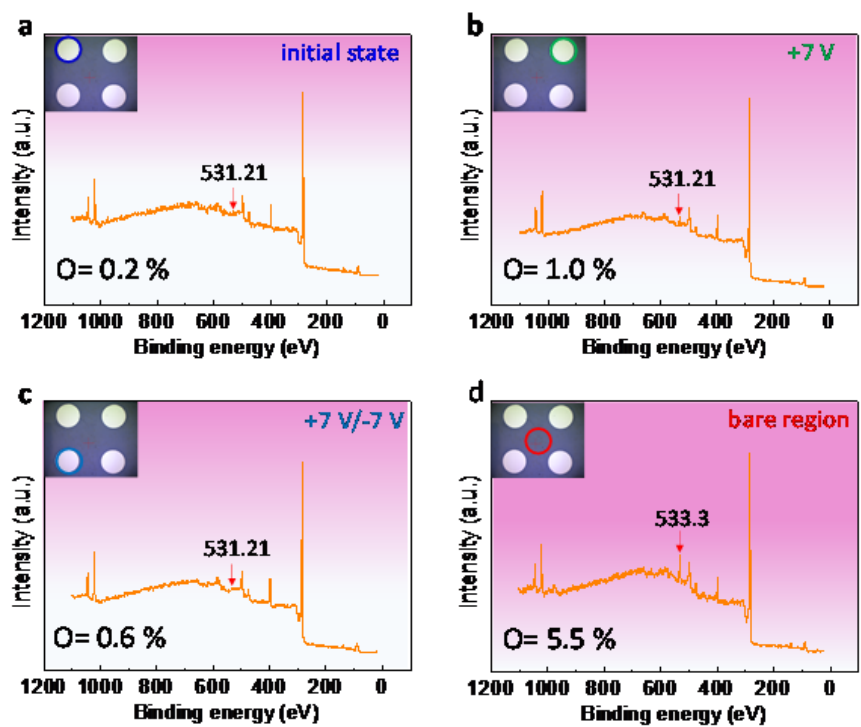

**Fig. S13| XPS survey spectra of ZnTPP films.** The measured regions are labeled in each inset. All spectrums are calibrated by align C 1s to 284.6 eV. Obviously, oxygen content increased markedly from 0.2% (**a**) to 1.0% (**b**) after polarized at +7V, and decreased to 0.6% (**c**) after negative polarized owing to a fraction of $O^{2-}$ reenter into $Al_2O_{3-x}$. However, the oxygen content very high (5.5%) in the bare region (**c**), which we ascribe to the adsorption of water in ambient atmosphere.

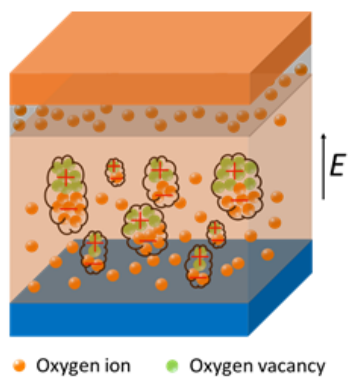

**Fig. S14|** A schematic illustration of the polarization of ZnTPP domains under an applied weak electrical field (V = 4 V).

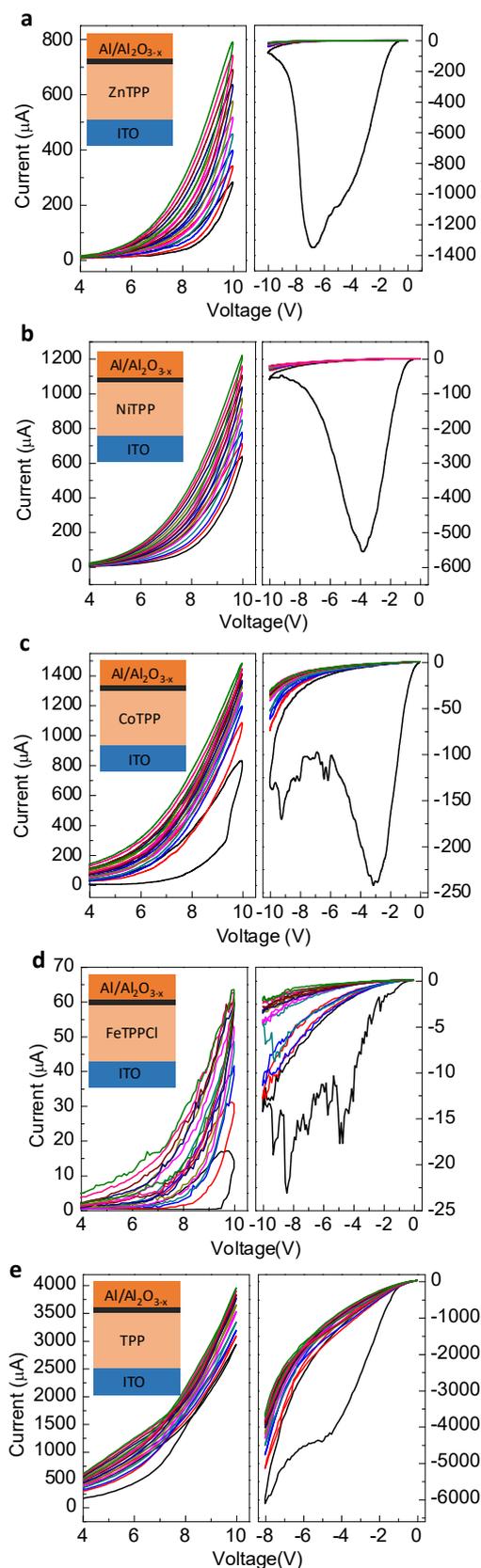

**Fig. S15| a-e**, Variable hysteretic *I-V* loops of MTPP thin films with identical configuration, ZnTPP (**a**) NiTPP (**b**), CoTPP (**c**), FeTPPCl (**d**) and TPP (**e**) measured at 10 V with voltage-sweep rates of 0.1 Vs$^{-1}$. Each of the device schematic is shown in the inset.

|  | Metal-O bond length (Å) | $\Delta E_O$ (eV) |
|---|---|---|
| ZnTPP | 1.63 | -1.866 |
| NiTPP | 1.86 | -1.110 |
| CoTPP | 1.92 | -0.266 |

**Table. S2|** The metal-oxygen bond length, binding energies of $O^{2-}$ on ZnTPP, NiTPP and CoTPP molecules, respectively.

a

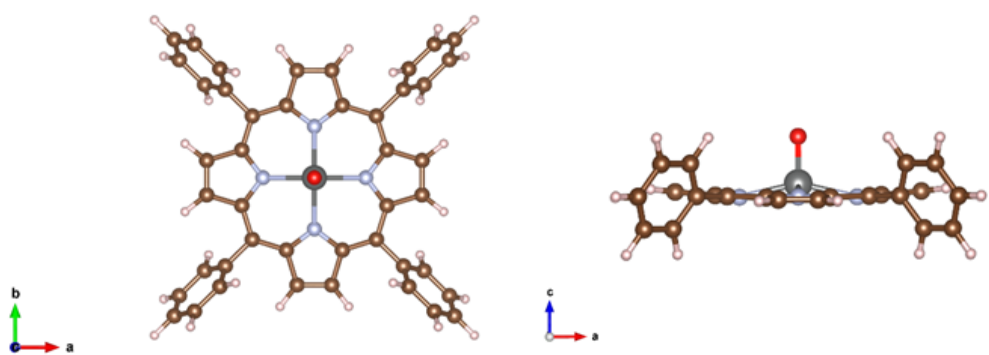

b

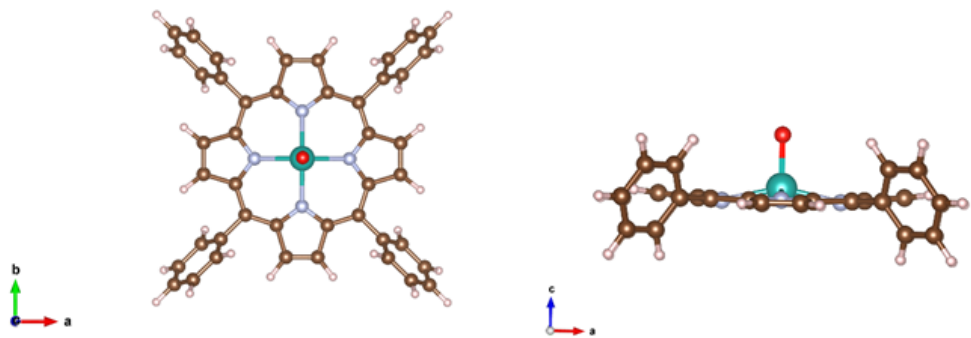

c

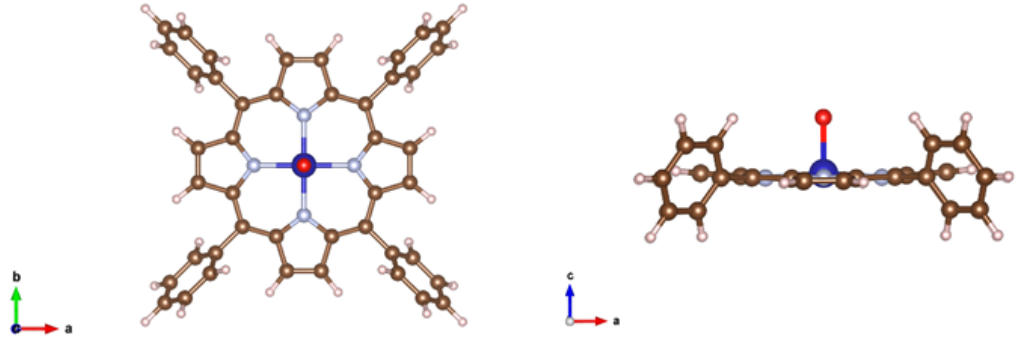

**Fig. S16|** The relaxed molecular structures of ZnTPP, NiTPP and CoTPP with oxygen adsorption. Gray, Cyan, and Blue atoms represent Zn, Ni, Co atoms in (**a-c**) respectively, red for oxygen, brown for carbon and light pink for hydrogen.

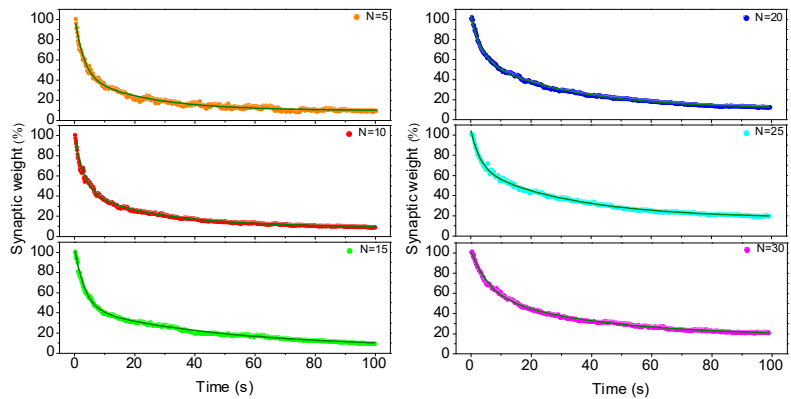

**Fig. S17|** Memory retention data (dots) and fitted (solid lines) recorded after different numbers of identical voltage pulse stimulations. The data are scaled by a prefactor $I_0$.

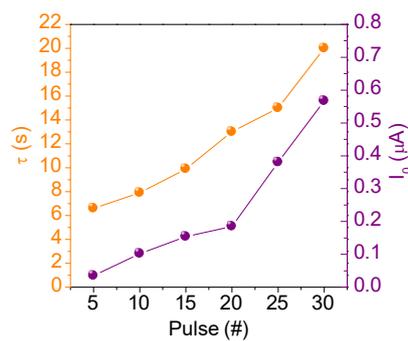

**Fig. S18|** Characteristic relaxation time ($\tau$) obtained through the fitting in panel Fig. S20 and the prefactor ($I_0$) plotted with respect to the number of stimulations (N).

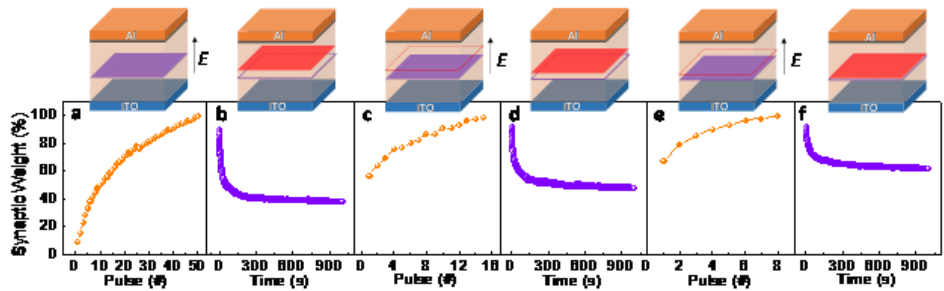

**Fig. S19| Demonstration of the 'learning–forgetting–relearning' process.** The insets show the corresponding migration conduction front model.

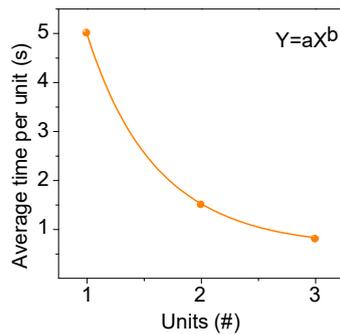

**Fig. S20| Wright's Cumulative Average Model.** In Wright's Model, the learning curve function is defined as $Y = aX^b$, where: Y = the cumulative average time (or cost) per unit, which is obtained by extracting data from Fig. S19; X = the cumulative number of units produced. a = time (or cost) required to produce the first unit (5 s in this model); b = slope of the function when plotted on log-log paper = log of the learning rate/log of 2.

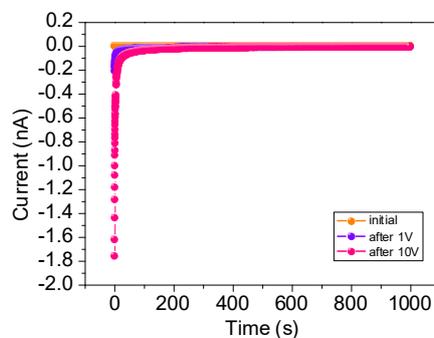

**Fig. S21|** Plots of the reverse current values after a series of 50 positive pulses (1 V, 500 ms and 10 V, 500 ms) were applied to the device. The current value was recorded by a read voltage of 0 V.

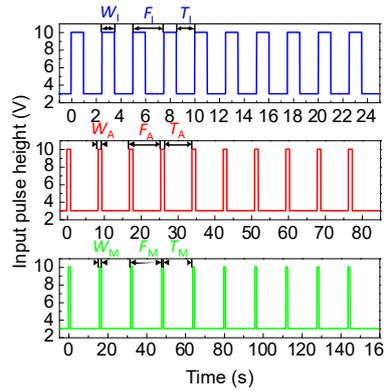

**Fig. S22| Spiking intervals applied to the device.** Images of the letters 'I', 'A', 'M' were continuous storing three times using ten identical amplitude (V = 10 V) and width ($W_I = W_A = W_M$ = 100 ms) inputs with intervals of $T_I$ = 1.5 s, $T_A$ = 7.5 s and $T_M$ = 15 s, respectively.

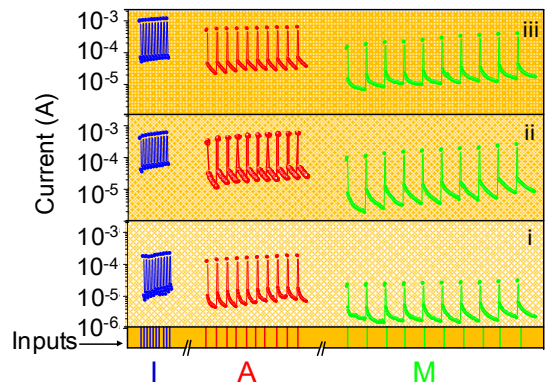

**Fig. S23|** Representative change in the conductance of an individual synapse response to various intervals and stored periods. The x axis shows the input pulse sequences with identical amplitude correspond to Fig. S25.

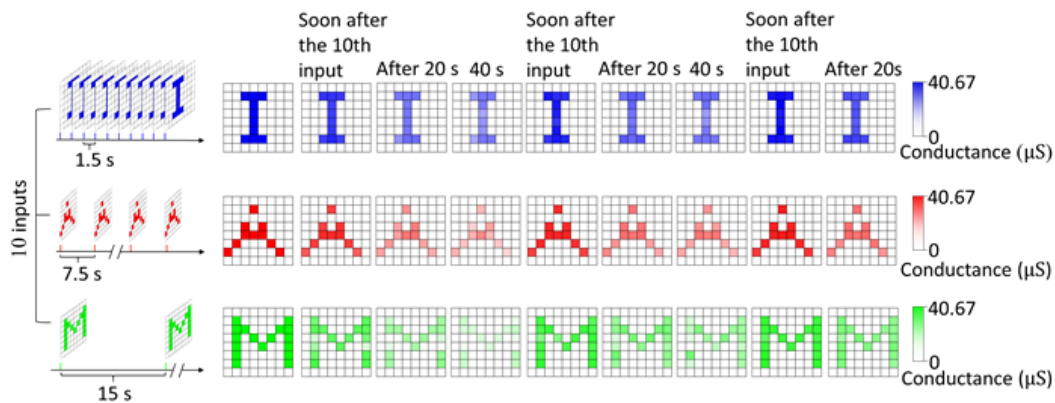

**Fig. S24| Visualizations of memorizing into an organic synapse array.** The letter 'M' was observed to persist slightly after 20 s and barely after 40 s from the last input, corresponding to STM mode. By contrast, the letters 'I' and 'A' was transferred to long-term memory (LTM) mode, which conform to the SRDP rules. Subsequent store

processes not only transfer STM to LTM (refer to the image of the letter 'M'), but also consolidate LTM mode remarkably(refer to the images of the letters 'I' and 'A'). The change in conductance at each pixel corresponds to the result for a single synapse, and the respective conductance profiles in the pixels are shown in **Fig. S26**.

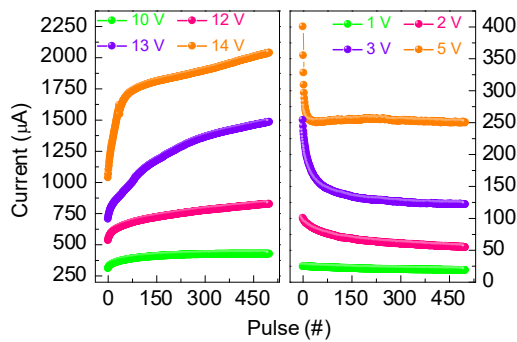

**Fig. S25| Voltage-dependent feature of ZnTPP devices.** Under the high voltage stimuli (V = 10 V, 12 V, 13 V, 14 V), the current increases more significantly as the voltage grows. In contrast, under low voltage stimuli (V = 1 V, 2 V, 3 V, 5 V), the current decrease and more obvious under higher voltage stimuli.

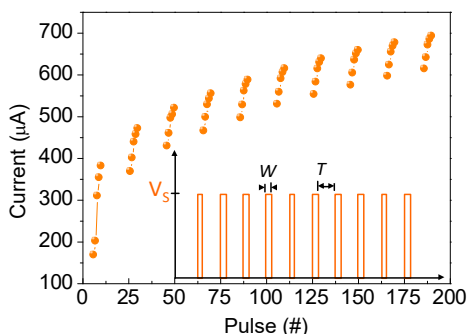

**Fig. S26|** Device current response to multiple subsequent voltage pulses (N = 10, V = 10 V, W = 500 ms, T = 1 s) which are employed to emulate the memorizing processes. The inset shows the shape of the applied voltage pulses.

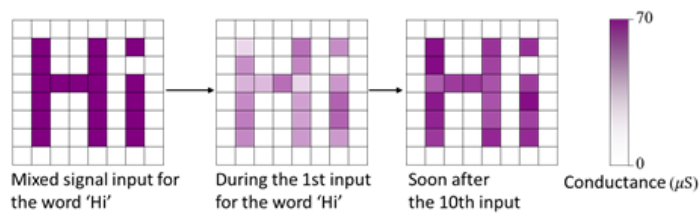

**Fig. S27|** Visualizations of the memorizing processes. The signals of word 'Hi' are continuously input into the memristor pixels with the voltage pulses shown in Fig. S30. Subsequently, the impression level of the word 'Hi' is enhanced gradually after ten input.

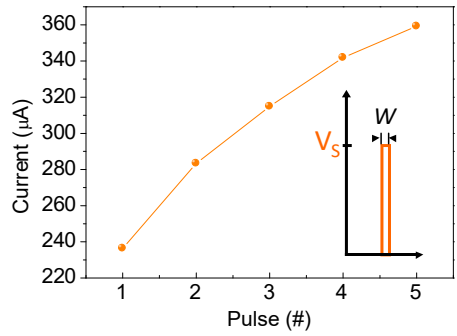

**Fig. S28|** Device current response to one voltage pulses (V = 10 V, W = 500 ms) which are employed to active the residual background information. The inset shows the shape of the applied voltage pulse.

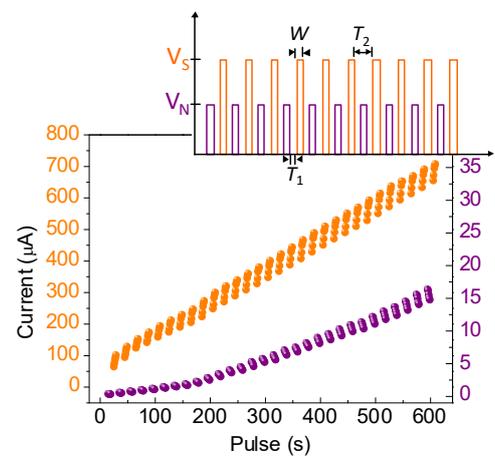

**Fig. S29|** Device current response to multiple subsequent voltage pulses which are employed to input the word 'Hi' in the activity-dependent information filtering process. The inset shows the shape of the applied voltage pulses, which consists of two segments: a noise pulse signal with low voltage (V = 5 V, W = 500 ms) and a useful pulse signal with high voltage (V = 10 V, W = 500 ms). $T_1$ = 500 ms, $T_2$ =1 S.

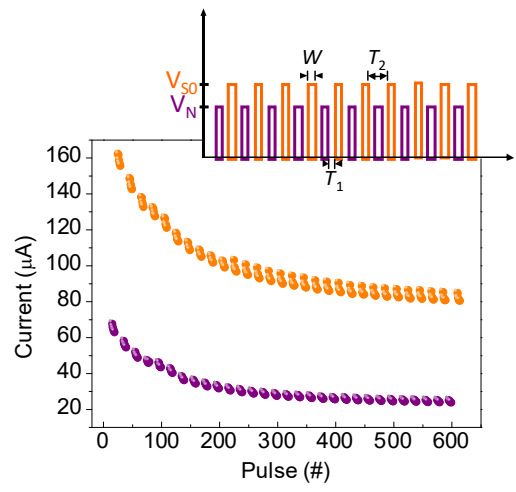

**Fig. S30|** Device current response to multiple subsequent voltage pulses which are employed to emulate the surrounding background information in Fig. 4b. The inset shows the shape of the applied voltage pulses, which consists of two segments: a noise

pulse signal with 6 V/500 ms, and an inhibition pulse signal with 5 V/500 ms. $T_1$ = 500 ms, $T_2$ =1 S.

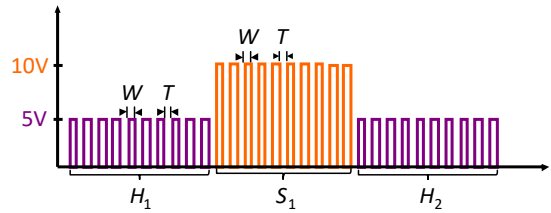

**Fig. S31|** Shematic illustration of alternant spiking mode for the emulations of habituation and sensitization including inhibitory stage $H$, excitatory stage $S$ ($W_I$ = $W_A$ = $W_M$ =100 ms, $T_I$ =1.5 s, $T_A$ = 7.5 s and $T_M$ = 15 s). The x axis is time and the y axis shows the input pulse sequences.

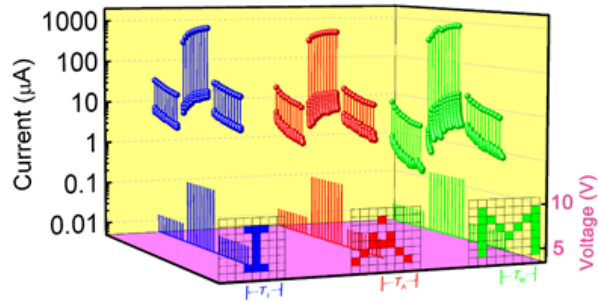

**Fig. S32|** The variation of the devices conductivity response to consecutive alternate modest stimuli (V = 5 V and $V_{read}$ = 3 V) and intense stimuli (10 V and 3 V) with different intervals ($T_I$ = 1.5 s, $T_A$ = 7.5 s, $T_M$ = 15 s).

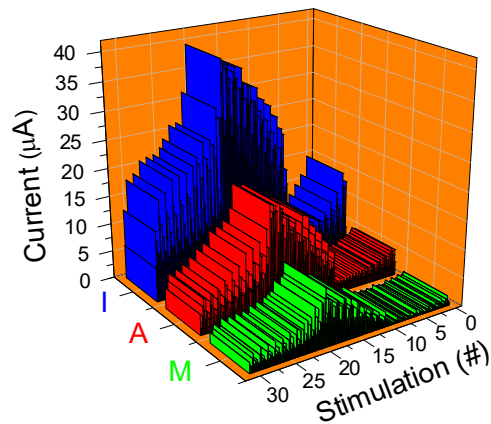

**Fig. S33|** Sensitization and habituation trends of ZnTPP synapse array response to various intervals and stored periods as a 3D plot. The sensitization level gets higher under higher-frequency stimuli and more periods, which cause progressive comprehensive amplification of our response to external activations.

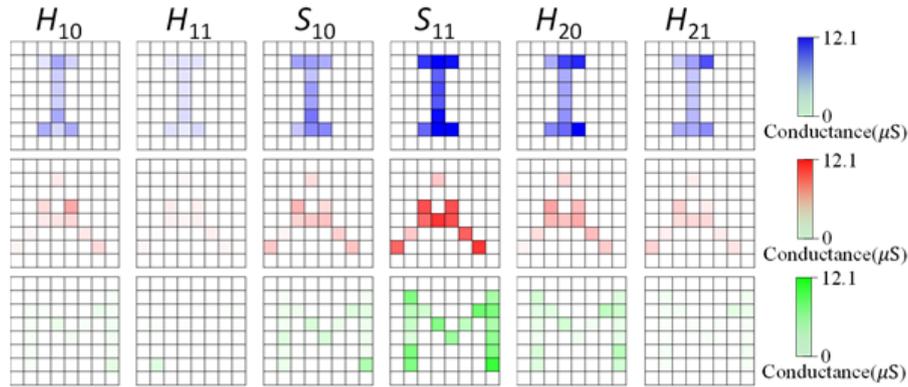

**Fig. S34|** Visualizations of sensitization and habituation behaviors. Two images in each habituation and sensitization stages represent the initial state and final state to reflect the overall conductive level as presented in Fig. S33.